\documentclass[lettersize,journal]{IEEEtran}
\usepackage{amsmath,amsfonts}
\usepackage{algorithmic}
\usepackage{algorithm}
\usepackage{array}
\usepackage[caption=false,font=normalsize,labelfont=sf,textfont=sf]{subfig}
\usepackage{textcomp}
\usepackage{stfloats}
\usepackage{url}
\usepackage{verbatim}
\usepackage{graphicx}
\usepackage{cite}
\usepackage{multirow} 
\usepackage{xcolor}
\usepackage{bbm}
\usepackage{amssymb}
\usepackage{makecell}
\usepackage{hyperref} 
\begin{document}

\title{Recognizing Ornaments in Vocal Indian Art Music with Active Annotation}

\author{Sumit Kumar$^{\dag}$, Parampreet Singh$^{\dag}$, Vipul Arora$^{\S}$\\
        $^{\dag}$Graduate Student Member, IEEE,\quad
        $^{\S}$Member, IEEE\\
        $^{\dag}$$^{\S}$Indian Institute of Technology, Kanpur
}

\markboth{Journal of \LaTeX\ Class Files,~Vol.~14, No.~8, August~2021}%
{Shell \MakeLowercase{\textit{et al.}}: A Sample Article Using IEEEtran.cls for IEEE Journals}


\maketitle

\begin{abstract}
Ornamentations, embellishments, or microtonal inflections are essential to melodic expression across many musical traditions, adding depth, nuance, and emotional impact to performances.
Recognizing ornamentations in singing voices is key to MIR, with potential applications in music pedagogy, singer identification, genre classification, and controlled singing voice generation. 
However, the lack of annotated datasets and specialized modeling approaches remains a major obstacle for progress in this research area.
In this work, we introduce R\=aga Ornamentation Detection (ROD), a novel dataset comprising Indian classical music recordings curated by expert musicians.
The dataset is annotated
using a custom Human-in-the-Loop tool for six vocal ornaments marked as event-based labels.
Using this dataset, we develop an ornamentation detection model based on deep time-series analysis, preserving ornament boundaries during the chunking of long audio recordings.
We conduct experiments using different train-test configurations within the ROD dataset and also evaluate our approach on a separate, manually annotated dataset of Indian classical concert recordings.
Our experimental results support the superior performance of our proposed approach over the baseline CRNN.  
\end{abstract}

\begin{IEEEkeywords}
Ornamentation detection, Singing techniques, MIR, Temporal Convolutional Networks, Indian Art Music
\end{IEEEkeywords}

\section{Introduction}

\IEEEPARstart{M}{usical}
ornaments are intentional variations around notes that bring expressiveness and complexity to a melody \cite{giraldo2018machine, 10.1007/978-981-97-1549-7_10}.
They are essential elements across musical traditions, shaping the music’s identity and emotional depth.
Automatic detection of ornaments in any music piece has broad applications in music education, performance analysis, and computational musicology, yet it remains significantly understudied, with very few works like~\cite{jpop, primadnn, pratyush_mtg}.

The task of ornamentation detection is typically formulated as an Audio Event Detection (AED) problem \cite{jpop, yamamoto2022deformable}, and therefore, several AED strategies and baseline models such as CRNN \cite{cakir2017convolutional, kao2018r} are commonly employed. However, this direct application presents certain challenges. 
Unlike standard audio events (e.g., clapping, car honks) that remain identifiable even when partially captured, musical ornaments
require their full temporal structure for accurate classification.
The common practice of segmenting audio into fixed-length chunks during pre-processing can fragment the ornaments across boundaries.
For instance, an ornament split between two chunks with one half at the end of one clip and the other at the beginning of the next would lose its structural integrity, leading to fragmented labels and misclassification during training.
Furthermore, effective progress in this area requires carefully curated datasets, specifically annotated to mark precise event boundaries of ornaments present within a musical piece. 
This level of annotation requires time-intensive efforts, domain expertise, and trained musicians, along with a structured annotation platform, making it an expensive, labor-intensive, and non-trivial task to undertake.

Analysis and detection of vocal ornamentations are highly relevant for both academic research and practical applications. One of the primary applications is in music pedagogy\cite{m3, merchan2024artificial, holland2013artificial}, 
where ornamentation detection can assess a student’s accuracy to replicate expressive nuances and provide focused feedback to improve vocal performance.
In controlled singing voice generation \cite{sintechsvs, luo2020singing}, ornamentation detection enables synthesis models to replicate expert singing styles more naturally by conditioning on detected ornamentation patterns. 
For singer identification, analyzing a singer’s distinctive use of ornamentations can enable recognition systems to differentiate vocalists based on their unique stylistic patterns \cite{loni2019robust}.

 \begin{figure*}[!th]
    \centering
    \includegraphics[scale=0.46]{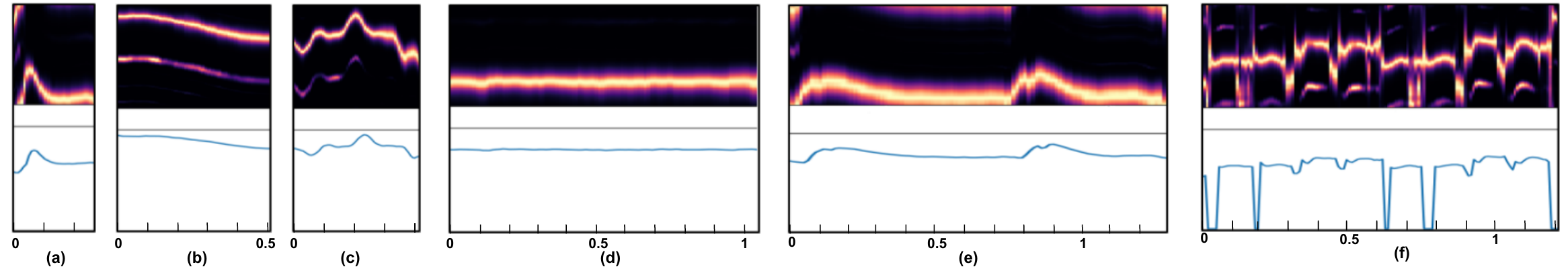}
    \caption{Chromagram (top) and Pitch Contour (bottom) representations of ornaments in IAM (a): \textit{Ka\d n},  (b): \textit{M\=ind}, (c): \textit{Murk\=i}, (d): \textit{Ny\=as svar}, (e): \textit{Andolan}, and (f): \textit{Gamak} in ROD dataset. x-axis represents time (in sec.)}
    \label{fig:ornament_classes}
    \vspace{-3mm}
    \end{figure*}

In this work, we present a comprehensive study on ornamentation detection in Indian Art Music (IAM). We introduce the Raga Ornamentation Detection (ROD) dataset, specifically curated for detecting ornaments in vocal performances, with detailed frame-level annotations for six distinct ornamentation types. To streamline the annotation process, we design a custom annotation tool incorporating a human-in-the-loop (HITL) framework 
within an active learning setting.
To address label fragmentation during chunking, we propose a don't care labelling-based training strategy. Using the ROD dataset, we train a deep learning model based on Temporal Convolutional Networks (TCNs) and demonstrate its performance over the baseline CRNN model. 
We experiment with various train-test split scenarios across different \textit{r\=ag\=as} and singers to assess the model’s robustness, and further evaluate its generalizability using a separate dataset of real-world IAM concert performances taken from Prasar Bharati\footnote{Prasar Bharati is India’s public broadcasting agency, comprising Doordarshan Television Network and All India Radio. It maintains an extensive archive of Indian classical music recordings.} Archives~\cite{param}.
We then conduct ablation studies on our TCN model, highlighting the importance of `don't care' labelling, periodic padding, and dilated convolutions on our proposed model. 
Our codes, dataset, pre-trained models, and demos are available on 
\href{https://github.com/madhavlab/2024_ornamentation}{https://github.com/madhavlab/2024\_ornamentation}.

\section{Ornaments in Indian Art Music}
Derived from the \textit{Sa\.msk\d{r}ita} word meaning ``beautification”, \textit{Ala\.nk\=aras} or ornaments constitute any pattern of musical decoration created by musicians within or across tones. 
In IAM, ornaments are essential to accurate \textit{r\=aga} rendition, whereas in Western music, they are often optional embellishments \cite{pratyush_mtg}. 
The six significant ornaments in IAM can be seen in Figure~\ref{fig:ornament_classes} are:
\subsubsection{\textit{Ka\d n} (grace notes)} \textit{Ka\d n} or the grace note
is a subtle, ornamental note in HCM that briefly touches the main note, either from above or below, without full articulation. It is rendered swiftly and delicately, enhancing melodic fluidity and expressiveness.

    \subsubsection{M\=i\.nd (gliding between notes)} In \textit{M\=i\.nd}, or glide, the pitch smoothly transitions between two notes, creating a continuous, flowing effect. Unlike \textit{Ka\d n}, which is a brief and subtle touch of an adjacent note, \textit{M\=i\.nd} involves a sustained and gradual pitch movement, often spanning multiple notes. \textit{M\=i\.nd} can be thought of as an elongated and more fluid counterpart to \textit{Ka\d n}.

    \subsubsection{Murk\=i (subtle renditions of note clusters)} A \textit{Murk\=i} is similar to a trill, typically involving the rapid and light alternation of two or three neighbouring notes. In fast-paced, folk-derived compositions, \textit{Murk\=is} are executed to sound light and sharp. In slower, more sensuous compositions, they are performed more smoothly, creating a languid rather than sharp effect.  Due to the flexibility inherent in its execution, \textit{Murk\=i} poses a significant challenge for automatic recognition and detection.
 
    \subsubsection{Ny\=as svar (emphasized resting notes)}\textit{Ny\=as svar,} or holding notes are specific notes within a \textit{r\=aga} that can be sustained for extended durations. \textit{Ny\=as svar} is the note on which a \textit{r\=aga} concludes.

    \subsubsection{Andolan (gentle oscillations)} An \textit{Andolan} is a gentle swing or oscillation that begins from a fixed note and reaches the periphery of another note. During these oscillations, it touches various microtones present between the notes. \textit{Andolan} can be conceptualized as multiple \textit{M\=i\.nd} in succession, punctuated by delicate holds that define its expressive motion.

    \subsubsection{Gamak (heavy oscillations)} A \textit{Gamak} is a rapid oscillation between two notes, performed with deliberate force and vigour. Compared to \textit{Murk\=i}, \textit{Gamak\=as} are generally more intense, faster, and executed with greater force, often lasting for a longer duration.

\section{Related Works}

\subsection{Datasets for ornamentation detection task}
Phonation Mode dataset \cite{15_juhan} 
consists of one-second recordings annotated at the clip level, covering four vocal modes, namely: neutral, pressed, breathy, and flow, produced by four different singers to detect phonation modes in sustained sung vowels. While it includes a broad pitch range, the discrete nature of the pitches lacks melodic context. Similarly, the PMSing dataset \cite{pmsing} extends phonation mode detection by incorporating adversarial discriminative training to improve classification accuracy. However, both datasets focus on phonation modes rather than detailed ornamentation.

VocalSet \cite{16_juhan} addresses this limitation by incorporating vocal performances in scales, arpeggios, long tones, and musical excerpts. It includes a diverse range of singing techniques, such as vibrato, trills, vocal fry, and inhaled singing. However, the dataset lacks strong labelling, as each audio file contains only one type of ornamentation, differing significantly from real-world music recordings. Moreover, several datasets have been annotated for singing voices in real-world settings \cite{17_juhan, svqdt, 18_juhan, jpop}. The KVT dataset \cite{17_juhan}, initially developed for singing voice tagging in K-POP, includes 70 vocal tags, with six related to singing techniques, such as whisper/quiet, vibrato, shouty, falsetto, speech-like, and non-breathy. However, it is not specifically designed for singing technique detection. The dataset includes annotations at both the song level and 10-second segment level, making it suitable for broad categorization but less detailed in ornamentation analysis. Similarly, the Singing Voice Query by Tag Dataset (SVQTD) \cite{svqdt} focuses on singing voice retrieval using descriptive tags, including timbral and phonation-based features. While it provides rich metadata for vocal characterizations, it lacks detailed temporal annotations, i.e., strong labels necessary for fine-grained ornamentation detection.

Further, the MVD dataset \cite{18_juhan}, designed for analyzing screams in heavy metal music, provides strong labelling for four types of screams—high fry, mid fry, low fry, and layered. Although it includes precise annotations, it was developed for scream detection rather than general singing techniques.
Furthermore, the authors in \cite{jpop} present a dataset that consists of 168 commercial J-POP songs, with annotations of various singing techniques marked with precise timestamps and corresponding vocal pitch contours. However, as the audio files are commercially licensed and only available through platforms such as Apple iTunes, the lack of open access poses challenges to reproducibility and may hinder further research.

Singing techniques are deeply rooted in the cultural and stylistic characteristics of a particular genre, making generalization across traditions challenging. Therefore, to address the task of ornamentation detection in IAM, we curate the ROD dataset, which we describe in detail in Section \ref{rod_dataset}.

\subsection{Analysis and Detection of Events}

The detection and analysis of events have been a long-standing area of research in audio and video signal processing \cite{mesaros2017dcase, weinland2011survey}. In the domain of audio event detection (AED), various methodologies have been proposed to automatically identify and classify acoustic events in diverse contexts, ranging from environmental sounds \cite{esc_1, esc_2, esc_3, bdcca}, to structured musical content like percussion stroke detection, drum transcription, etc.\cite{stroke_1, drum}. Similarly, in video signal processing, action segmentation techniques aim to segment and recognize temporal sequences of human activities using deep learning-based architectures such as Temporal Convolutional Networks (TCNs) \cite{ding2023temporal, lea2016temporal, edtcn}.

Given the conceptual overlap between general event detection tasks and ornamentation detection in music, we approach the latter as a specialized case of AED. Specifically, our work builds upon prior research in AED and extends it to the musical domain. The baseline study for this task \cite{jpop} adopts a conventional Convolutional Recurrent Neural Network (CRNN) architecture, formulating ornamentation detection as a frame-wise classification problem. In contrast, we employ a Temporal Convolutional Network (TCN)-based approach, which is discussed in detail in Section \ref{edtcn_model}.

\section{Dataset Description}
\label{rod_dataset}
The Rāga Ornamentation Detection (ROD) dataset is a curated collection of Hindustani Classical Music (HCM) vocal performances for the study of ornamentations in IAM. It consists of 212 audio files recorded by two expert singers, totaling 4.08 hours of audio. Singer 1 and singer 2 contributed 108 and 104 recordings, respectively. Each audio file is a mono-channel WAV file recorded at 44.1 kHz sampling rate and 32-bit precision, using an Audio-Technica AT2020 condenser microphone and a laptop with a 12-core Intel Core i5-12500H processor. 
The dataset spans four distinct \textit{r\=agas} across the two singers, with details summarized in Table~\ref{dataset_stats}. While recording, the singers were given the flexibility to design structured lessons, maintaining that all exercises within a lesson adhered to a single \textit{r\=aga}. Lessons were organized to progress from easy to difficult levels, targeting pedagogical objectives.
During recordings, tanpura and tabla accompaniments were provided; however, vocals were recorded separately as mono tracks to ensure clean vocal capture. Metadata, including tonic, \textit{tāla}, tempo (BPM), and accompaniment files, is provided to allow later merging of accompaniment and vocal tracks if needed.

\subsection{Annotation Process} 

The annotation process follows a two-stage approach by two sets of expert annotators. 
In the first stage, the first set of annotators manually annotate their respective recordings by listening to the audio files. Using Audacity~\cite{audacity} software, they mark the onset (start time), offset (end time), and the type of ornamentation (class) for each occurrence. 
The ornaments are strongly labelled as non-overlapping events, similar to those in Audio Event Detection (AED) tasks.
Since labelling musical ornamentations differs significantly from labelling typical audio events in AED tasks \cite{SED, bdcca}, subjectivity and stylistic variations between annotators can introduce discrepancies in the first round of annotation. annotators may mark different labels for similar patterns or merge two closely occurring ornamentations into a single label, leading to variations in event segmentation.

To refine the annotations, a second stage is carried out by another set of annotators using a model trained on the first-round labels. These annotatorsare given the same audio recordings along with a Tkinter-based UI that displays chromagram visualizations, pitch contours, first-stage annotations, and the model’s predictions. This setup, in an active learning framework, allows them to make precise adjustments to event timings.
The second round of annotation addresses inconsistencies from the first round by refining event boundaries and ensuring that closely occurring ornamentations are distinctly labelled. For instance, in cases where multiple \textit{Ka\d nsvar} appear in quick succession, they are individually marked rather than being merged into a single event.

\begin{figure*}
    \centering
    \includegraphics[scale=0.35]{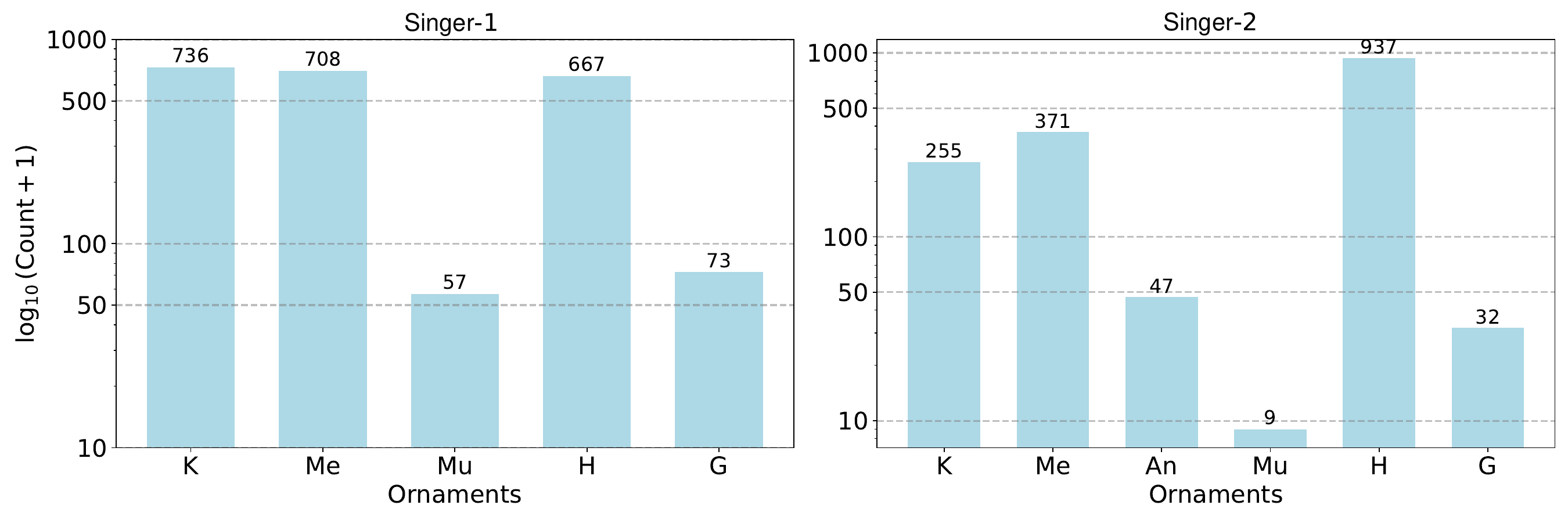}
    \caption{Class-wise ornament frequency for Singer 1 (left) and Singer 2 (right), with ornament classes on the x-axis and log-scaled counts on the y-axis. K, Me, An, Mu, H, and G represent \textit{Ka\d n}, \textit{M\=ind}, \textit{Andolan}, \textit{Murki}, \textit{Ny\=as Svar}, and \textit{Gamak} respectively.}
    \label{fig: class_distribution}
    \vspace{-3mm}
    \end{figure*}

Figure \ref{fig: class_distribution} shows ornamentation frequency observed in the performances of two vocal singers. 
Singer-1 demonstrates higher usage across most categories. This variation can be attributed to differences in their musical \textit{ghar\=an\=as}\cite{harrison_gharana} (lineage-based musical tradition of IAM), which influences ornamentation and singing styles in general.

\subsubsection{Ornamentation-Specific Annotation Rules}To standardize the labelling process, specific duration constraints and clear event boundaries are established for each ornamentation type. For \textit{Ka\d nsvar}, a maximum duration is limited to 0.35 seconds to avoid confusion with \textit{M\=ind}. In the case of \textit{M\=ind}, a minimum duration of 0.45 seconds is required, and 
only the gliding portion is labelled, excluding sustained holds after the glide. 
For \textit{Ny\=as Svar}, a minimum duration of 0.6 seconds is set without imposing an upper limit. \textit{Murk\=i} is constrained to a duration range of 0.4 to 1.0 seconds.
For \textit{Andolan}, a minimum duration of 1.0 seconds is required to distinguish it from isolated \textit{M\=ind}. Lastly, \textit{Gamak} is annotated only if it persisted for at least 0.7 seconds to avoid misclassification with 
\textit{Ka\d nsvar} or \textit{Murk\=i}.

\subsection{Inter-Annotator Agreement}
To account for subjectivity that might be seen in the annotation because of differences in expertise, perception, stylistic variations, and interpretations across \textit{Ghar\=an\=as}, we assess inter-annotator agreement (IAA) using Cohen’s Kappa score. 
An additional set of annotators, using the rules and annotation setup, independently perform a blind round of annotation for both datasets, resulting in a Cohen’s Kappa score of 0.79. 

\begin{table}[htbp]
    \centering
    \caption{Duration-wise distribution of audio recordings across different rāgas and singers in the ROD dataset.}
    \begin{tabular}{|c|c|c|}
        \hline
        \textbf{Singer} & \textbf{\textit{R\=aga}} & \textbf{Duration (in seconds)} \\
        \hline
        \multirow{2}{*}{Singer 1} & \textit{B\=ageshr\=i} & 2188 \\
        \cline{2-3}
        & \textit{Bh\=up\=ali} & 1911 \\
        \hline
        \multirow{3}{*}{Singer 2} & \textit{Bhairav} & 3345 \\
        \cline{2-3}
        & \textit{Bh\=up\=ali} & 6967 \\
        \cline{2-3}
        & \textit{Darb\=ar\=i} & 282 \\
        \hline
    \end{tabular}
    
    \label{dataset_stats}
\end{table}
\begin{figure}[t]
\centering
\includegraphics[scale=0.43]{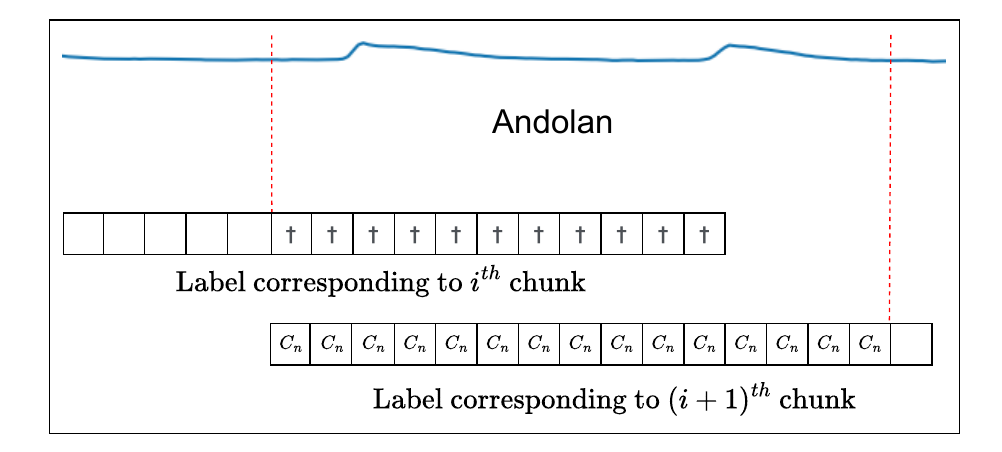}
\caption{An illustration of audio chunking using overlapping windows and corresponding label encoding. Label $C_n$ corresponds to class \textit{Andolan} (within red lines) and $\dagger$ is the `don't care' label.}
\label{fig:chunking}
\vspace{-3mm}
\end{figure}
\section{Proposed Methodology}
\begin{figure*}[t]
    \centering
    \includegraphics[scale=0.7]{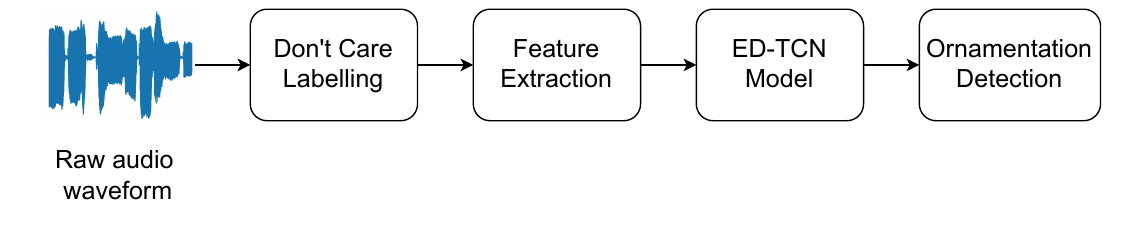}
    \caption{An overview of the proposed methodology for the ornamentation detection task}
    \label{overview}
    \vspace{-3mm}
    \end{figure*}

We formulate the ornamentation detection task in analogy with audio event detection (AED), while introducing critical adaptations tailored to the unique characteristics of ornamentations. The proposed methodology comprises several stages, including data pre-processing, feature extraction, and the design of a deep learning model as shown in Figure \ref{overview}. Unlike conventional AED approaches, our pipeline incorporates modifications to preserve the temporal structure of ornamentations during both pre-processing and model training, ensuring accurate temporal localization and characterization of ornaments.

\subsection{Data Preprocessing}
\label{preprocess}

In general, audio recordings in a dataset vary in length \cite{fonseca2020fsd50k, panayotov2015librispeech, poliner2005classification}. To use audio or extracted features as input to a deep learning model, the recordings are generally segmented into fixed-duration chunks \cite{jpop}. However, in our task, this chunking strategy can adversely affect ornaments occurring at the boundaries of segments. For instance, an \textit{Andolan} that is abruptly segmented may resemble a \textit{M\=ind} or a short portion of \textit{Gamak} may resemble \textit{Murk\=i}. To address this issue, we propose a novel chunking technique that preserves the integrity of ornaments at segment boundaries, as detailed below.

Given the dataset $\mathcal{D} = \{(x_i, \{(o_{ij}, f_{ij}, c_{ij})\}_{j=1}^{N_i})\}_{i=1}^M$, where $x_i$ denotes the $i$-th audio file, and each event within $x_i$ represented by its onset $o_{ij}$, offset $f_{ij}$, and class label $c_{ij}$, we propose a chunking strategy for the ornamentation detection task. Here, $M$ is the total number of audio files in the dataset, and $N_i$ is the number of ornaments in the $i^{\text{th}}$ file. The objective is to create chunks of length $t$ seconds while ensuring that events are not abruptly truncated at chunk boundaries.

For each audio file $x_i$, the chunking process begins at time $t_{i0} = 0$ seconds. Each chunk, denoted as $s_{ik}$, spans the interval $[t_{ik}, t_{ik} + t]$, where $t$ and $k$ denote the duration and index of the chunk corresponding to the $i^{\text{th}}$ audio file $x_i$, respectively. In our proposed chunking method, if an event $e_{ij} = (o_{ij}, f_{ij}, c_{ij})$ starts within the interval $[t_{ik}, t_{ik} + t]$ but ends after $t_{ik} + t$, the next chunk $s_{i(k+1)}$ will start from the onset $o_{ij}$ of this event. Mathematically, if $o_{ij} \leq t_{ik} + t < f_{ij}$, then $t_{i(k+1)} = o_{ij}$.

To ensure that truncated events do not participate in model training, we label the truncated portion as `don't care' (denoted as $\dagger$). Specifically, for the $k$-th chunk $s_{ik}$, if $t_{ik} \leq o_{ij} < t_{ik} + t$ and $t_{ik} + t < f_{ij}$, then the segment $[o_{ij}, t_{ik} + t]$ within the $k$-th chunk is labelled as a don't care label, i.e., $\dagger$. The subsequent chunk $s_{i(k+1)}$ will cover the interval $[o_{ij}, o_{ij} + t]$, and the segment $[o_{ij}, f_{ij}]$ within this chunk will be labelled with the original class label $c_{ij}$. The proposed chunking technique is illustrated in Figure \ref{fig:chunking}.

Formally, let $\mathcal{S}_i$ be the set of chunks for the audio file $x_i$, and $\mathcal{L}_i$ be the corresponding labels:
\[
\mathcal{S}_i = \{s_{ik}\}_{k=0}^{K_i}, \quad \mathcal{L}_i = \{l_{ik}\}_{k=0}^{K_i}
\]
where $K_i$ represents the total count of chunks associated with the audio file $x_i$, and each chunk $s_{ik}$ is defined as:
\[
s_{ik} = [t_{ik}, t_{ik} + t]
\]
The labels $l_{ik}$ are assigned as follows:
\[
l_{ik} = \begin{cases} 
c_{ij}, & \text{if } o_{ij} \leq t_{ik} \text{ and } f_{ij} \leq t_{ik} + t \\
\dagger, & \text{if } o_{ij} \leq t_{ik} \text{ and } t_{ik} + t < f_{ij}
\end{cases}
\]

To further process the audio chunks, we extract chromagram features \cite{chroma}, denoted as $\mathcal{X} = \{X_k\}_{k=1}^L$, for all $L$ chunks across the dataset. Here, $L = \sum_{i=1}^M K_i$ denotes the total number of chunks across all $M$ audio files. Each $X_k \in \mathbb{R}^{F \times T}$, with $F$ denoting the number of frequency bins and $T$ representing the number of time frames. Let the corresponding labels be $\mathcal{Y} = \{Y_k\}_{k=1}^L$, where each $Y_k \in \mathbb{R}^T$ encodes the ground truth for each time frame of $X_k$. The set $(\mathcal{X}, \mathcal{Y})$ is used for training and testing the model.

\subsection{Model} \label{edtcn_model}
We introduce the Encoder-Decoder Temporal Convolutional Network (ED-TCN) model, inspired by Lea et al. \cite{edtcn} to our task, with chromagram features as input. The authors in \cite{edtcn} designed ED-TCN specifically for the task of action segmentation in video signals. Since our task is similar to action segmentation \cite{j_gall, j_gall2013} with respect to the temporal dependency of events, we customize the ED-TCN model to fit our task by introducing periodic padding \cite{PP} and dilated convolution inspired by the Wavenet\cite{wavenet} architecture.

The ED-TCN model comprises the following components:
\subsubsection{Encoder} 
\begin{figure}[t]
    \centering
    \includegraphics[width=1\columnwidth]{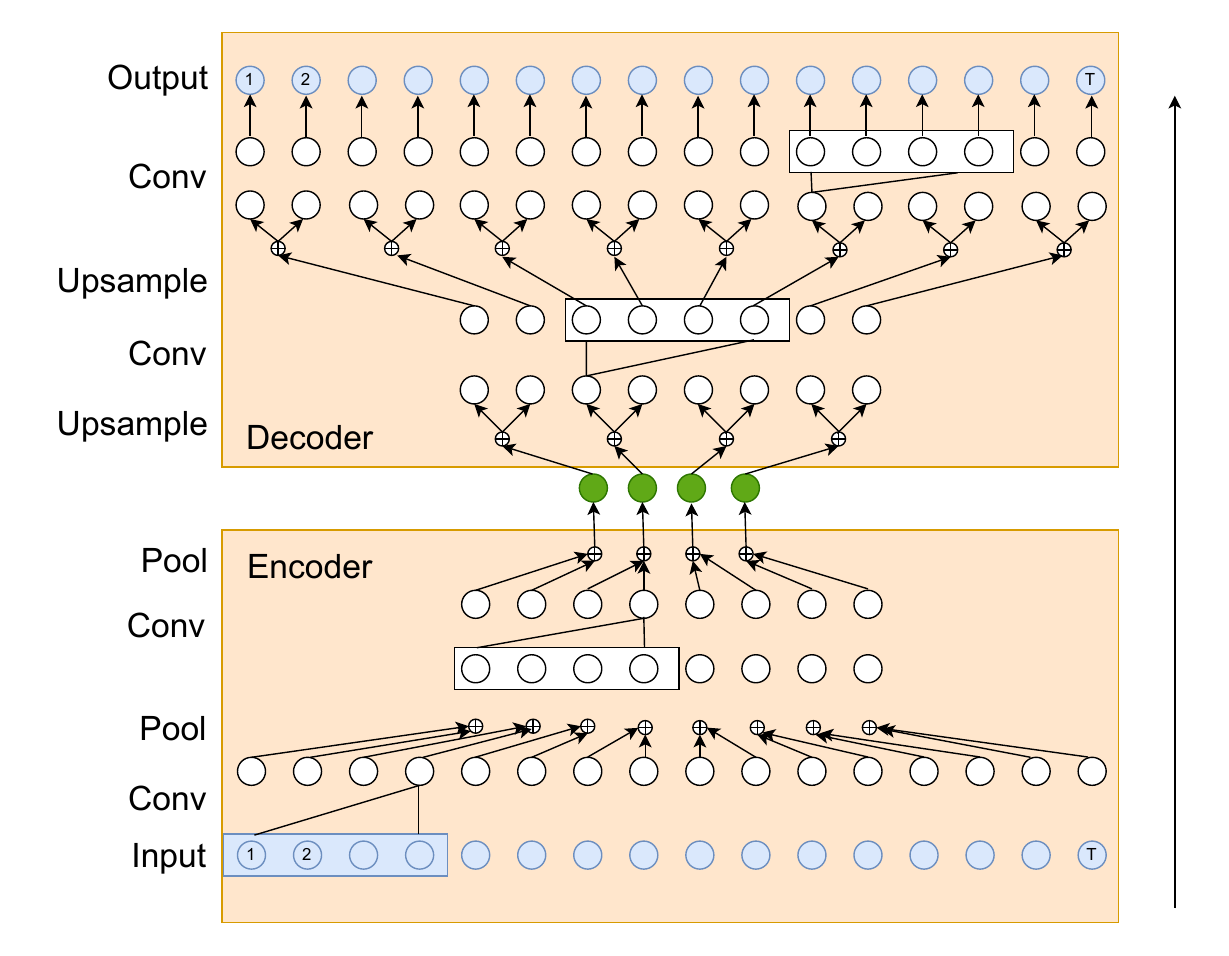}
    \caption{Encoder-Decoder Temporal Convolutional Network}
    \label{edtcnfig}
    \vspace{-3mm}
    \end{figure}

The encoder consists of $L$ layers denoted by $E^{(l)} \in \mathbb{R}^{F_l \times T_l}$, where $F_l$ is the number of filters applied for the convolution operation in $l^{th}$ layer and $T_l$ is the number of time steps in the corresponding layer. The encoder part of the network applies a series of dilated convolutions to the input chromagram features, $\mathcal{X}$ as shown in Figure \ref{edtcnfig}. These convolutions progressively downsample the temporal dimension such that $T_l = \frac{1}{2}T_{l-1}$ using max pooling with pool size of 2 time frames
Let $W = \left\{W^{(i)}\right\}_{i=1}^{F_l} \quad \text{for} \quad W^{(i)} \in \mathbb{R}^{d \times F_{l-1}}$ where $W$ is the set of convolutional filters and $d$ is the filter size. Let the corresponding bias vector be $ b \in \mathbb{R}^{F_l}$. Given the output from the previous layer, $E^{(l-1)}$, the $l^{th}$ layer is computed as
    \begin{equation}
    E^{(l)} = \sigma(W \ast E^{(l-1)} + b)
    \label{eq:label}
    \end{equation}
    where $\sigma$ is the non-linear activation function and $\ast$ denotes the convolution operation.

\subsubsection{Decoder} The architecture of the decoder is similar to that of the encoder. The pooling layer is simply replaced with an upsampling layer. Upsampling is utilized to ensure that the number of time frames in the input layer corresponds to those in the output layer. The decoder also consists of $L$ layers denoted by $D^{(l)} \in \mathbb{R}^{F_l \times T_l}$ where $l \in \{L, \ldots,   1\}$. 

\subsubsection{Dilated Convolutions} To effectively capture long-range temporal dependencies, we incorporate dilated convolutions into the encoder of the ED-TCN model. Unlike standard convolutions that process consecutive time steps, dilated convolutions apply the filter over input features with defined gaps, known as dilation rate, between elements. 

\subsubsection{Periodic Padding} To leverage the periodic nature of musical notes, we introduce a custom padding layer, \texttt{PeriodicPadding}, which replaces the standard zero padding used in convolution operations. This periodic padding ensures that the convolution operation respects the cyclic structure of the extracted chromagram features, improving the model's ability to detect repetitive patterns. Mathematically, the periodic padding operation can be described as:
      \[
       PeriodicPadding = 
       \begin{bmatrix}
       X_{\text{top}} \\
       X \\
       X_{\text{bottom}}
       \end{bmatrix}
       \]

       where $X$ is the chromagram feature matrix $\in \mathbb{R}^{F \times T}$, $X_{top} \in \mathbb{R}^{p \times T}$ is the set of last $p$ rows of the $X$, and $X_{bottom} \in \mathbb{R}^{p \times T}$ is the set of first $p$ rows of the $X$ such that $X_{top} = X[-p:]$ and $X_{bottom} = X[:, p]$ where $p$ is the padding size.

\subsubsection{Classification Layer} At the end of the decoder, we apply a Time-distributed dense layer with \(C\) neurons corresponding to the number of classes. This dense layer acts as the classifier, mapping the final feature map to the class probabilities for each time frame.  
    
    
    The classification layer applies a linear transformation followed by a softmax activation to produce the class probabilities such that:
    \begin{equation}
    \label{eq_output}
    \hat{Y}_t = \text{softmax}(U D_t^{(1)} + b)
\end{equation}

where \(U  \in \mathbb{R}^{C \times F_1}\) is the weight matrix, \(b \in \mathbb{R}^C\) is the bias vector, and $D_t^{(1)}$ represents the last layer of decoder, and \(\hat{Y}_t \in \mathbb{R}^C\) represents the predicted class probabilities for each time frame \(t \in \{1, 2, \ldots, T\}\).

Let the ED-TCN model be denoted by $f_{\theta}$ where $\theta$ are the model parameters. The parameters $\theta$ are updated using the gradient descent algorithm: $\theta \leftarrow \theta - \alpha \boldsymbol{\triangledown}_{\theta} L(f_{\theta}),$
where $\alpha \in \mathbb{R^+}$ is the learning rate and $L$ is the categorical cross-entropy loss defined as :

\begin{equation}
          L = -\frac{1}{\sum_{t=1}^{T} \mathbbm{1}_{\{Y_t \neq \dagger\}}} \sum_{t=1}^{T} \mathbbm{1}_{\{Y_t \neq \dagger\}} Y_t \log(\hat{Y}_t)
\end{equation}
where $Y_t \in \mathbb{R}^C$ denotes the one-hot representation of the ground truth label for $t^{th}$ time frame of the chromagram $X_t$ and $\hat{Y}_t = f_{\theta}(X_t)$ denotes the predicted probability distribution over classes as defined in \ref{eq_output}. The indicator function $ \mathbbm{1}_{\{Y_t \neq \dagger\}}$ denotes that while loss calculation, only those time-frames are considered whose labels are not `don't care' labels $\dagger$.

\section{Experiments} 

\begin{table}[]
    \centering
    \setlength{\tabcolsep}{1.5pt}  
    \caption{Overview of experimental data splits used to evaluate generalization across singers and rāgas. $S_1^{tr}$ and $S_2^{tr}$ denote the training subsets  of Singer~1 ($S_1$) and Singer~2 ($S_2$) data, respectively, while $S_1^{te}$ and $S_2^{te}$ represent the corresponding testing subsets.}
    \begin{tabular}{|c|l|l|l|}
        \hline
        \textbf{Exp} & \textbf{Training Data} & \textbf{Testing Data} & \textbf{Testing Type} \\ 
        \hline
        1 & $S_1^{tr}$ + $S_2^{tr}$ & $S_1^{te}$ + $S_2^{te}$ & General \\ 
        2 & $S_1^{tr}$ & $S_1^{te}$ & Intra-Singer \\ 
        3 & $S_2^{tr}$ & $S_2^{te}$ & Intra-Singer \\ 
        4 & $S_1$ & $S_2$ & Inter-Singer \\ 
        5 & $S_2$ & $S_1$ & Inter-Singer \\ 
        6 & $S_1$, Rāga Bageshree & $S_1$, Rāga Bhoopali & Rāga-Specific \\ 
        7 & $S_1$, Rāga Bhoopali & $S_1$, Rāga Bageshree & Rāga-Specific \\ 
        8 & $S_2$, Rāga Bhairav & $S_2$, Rāga Bhoopali & Rāga-Specific \\ 
        9 & $S_2$, Rāga Bhoopali & $S_2$, Rāga Bhairav & Rāga-Specific \\ 
        \hline
    \end{tabular}
    \label{tab:data_splitting}
    \vspace{-3mm}
\end{table}

Since the audio files have variable lengths, we segment them into 10-second chunks using the chunking technique discussed in Section \ref{preprocess}, followed by extracting chromagram features.
For computing chromagrams, we calculate the Short-time Fourier transform (STFT) of the audio chunks using a 4096-point FFT, a 35 ms sliding Hanning window with a hop-size of 17.5 ms
Since ornamentation detection relies heavily on capturing continuous pitch structures, we examine chromagrams with 12 and 120 bins.

\subsection{Baseline} We adopt the model proposed in \cite{jpop} as the baseline. The input to the model is chromagram features, $X \in \mathbb{R}^{F \times T}$ as mentioned above. We use three sets of 2D-Convolution, BatchNormalization, and Maxpooling layers. We use [32, 64, 128] kernels of size (3, 3). Posterior to it, we apply two bidirectional-GRU layers, each having 8 recurrent units. We use ReLU non-linearity after both convolutional and recurrent layers. Following the bi-GRU layers, we apply a Time-distributed dense layer with softmax activation having C = 6 neurons corresponding to six classes. The model is trained for 100 epochs using the Adam optimizer with a learning rate of 0.001.

\subsection{Proposed Model} The input to the proposed ED-TCN model is again chromagram $X \in \mathbb{R}^{F \times T}$ as mentioned above. The encoder consists of L = 4 1D-Convolutional layers having $W = $[32, 64, 128, 256] number of filters with ReLU activation at each layer. We also apply dilation \cite{wavenet} at each convolutional layer with a linearly increasing dilation rate [1, 2, 3, 4] at each layer. The decoder follows a similar architecture to the encoder. It consists of 1D-Convolutional layers having [256, 128, 64, 32] number of filters with ReLU activation at each layer.  The size of each 1D-Convolutional filter is 5. Dilation rates are applied in the opposite sequence to those in the encoder. A Spatial-Dropout layer with a dropout rate of 0.3 is used in both the encoder and decoder. A Time-Distributed dense layer with softmax having C = 6 neurons follows the decoder. The model is trained to 3000 epochs using the Adam optimizer with a learning rate of 0.001.

We test our model across different splits of the dataset, including general overall evaluation, intra-singer, inter-singer, and also Raga-specific evaluation, as shown in Table~\ref{tab:data_splitting}, detailed in section~\ref{sec:results}.
To further evaluate the generalizability of our model to real-world IAM performances, we fine-tune the best-performing model on the Prasar Bharati dataset~\cite{param}. 
In this setup, we first use DEMUCS~\cite{defossez2021hybrid} to extract the vocal stems from the polyphonic audio recordings. We then use our model, originally trained on the ROD dataset, as a fixed feature extractor and fine-tune only the final layers on the new, unseen dataset.

 \subsection{Evaluation Metrics}
We report Precision, Recall, and F1-score averaged over the classes for our task. Since it is practically infeasible to annotate the boundaries of events precisely, the model struggles to predict the exact boundary of the events. It is a common practice to use a collar around the event boundaries, to address this issue. The authors in \cite{SED, m3} use a 200 ms collar around the event boundaries during evaluation. We adopt similar collar-based evaluation metrics for our task.

\begin{table*}[t]
\centering
\caption{Performance comparison of the baseline and proposed model across nine evaluation experiments, reported with and without a 200 ms collar for ROD dataset.}
\begin{tabular}{|l|llllll|llllll|}
\hline
\multicolumn{1}{|c|}{\multirow{3}{*}{\textbf{Experiment}}} & \multicolumn{6}{c|}{\textbf{Baseline}}                                                                                                                                                 & \multicolumn{6}{c|}{\textbf{Proposed Model}}                                                                                                                                           \\ \cline{2-13} 
\multicolumn{1}{|c|}{}                                   & \multicolumn{3}{c|}{\textbf{Without collar}}                                                         & \multicolumn{3}{c|}{\textbf{With 200 ms collar}}                                & \multicolumn{3}{c|}{\textbf{Without collar}}                                                                  & \multicolumn{3}{c|}{\textbf{With 200 ms collar}}                                         \\ \cline{2-13}
\textbf{}                                                & \multicolumn{1}{l|}{\textbf{P}} & \multicolumn{1}{l|}{\textbf{R}} & \multicolumn{1}{l|}{\textbf{F1}} & \multicolumn{1}{l|}{\textbf{P}} & \multicolumn{1}{l|}{\textbf{R}} & \textbf{F1} & \multicolumn{1}{l|}{\textbf{P}} & \multicolumn{1}{l|}{\textbf{R}} & \multicolumn{1}{l|}{\textbf{F1}} & \multicolumn{1}{l|}{\textbf{P}} & \multicolumn{1}{l|}{\textbf{R}} & \textbf{F1} \\ \hline
\textbf{1}                                               & \multicolumn{1}{l|}{74.92}           & \multicolumn{1}{l|}{73.78}           & \multicolumn{1}{l|}{74.35}            & \multicolumn{1}{l|}{81.08}           & \multicolumn{1}{l|}{75.28}           &78.71             & \multicolumn{1}{l|}{\textbf{90.56}}      & \multicolumn{1}{l|}{\textbf{89.56}}      & \multicolumn{1}{l|}{\textbf{90.06}}       & \multicolumn{1}{l|}{\textbf{94.14}}      & \multicolumn{1}{l|}{\textbf{89.20}}      & \textbf{91.98}       \\ \hline
\textbf{2}                                               & \multicolumn{1}{l|}{65.91}           & \multicolumn{1}{l|}{64.86}           & \multicolumn{1}{l|}{65.38}            & \multicolumn{1}{l|}{71.95}           & \multicolumn{1}{l|}{66.83}           &69.30             & \multicolumn{1}{l|}{\textbf{90.15}}      & \multicolumn{1}{l|}{\textbf{89.04}}      & \multicolumn{1}{l|}{\textbf{89.59}}       & \multicolumn{1}{l|}{\textbf{93.54}}      & \multicolumn{1}{l|}{\textbf{89.40}}      & \textbf{91.42}       \\ \hline
\textbf{3}                                               & \multicolumn{1}{l|}{69.77}           & \multicolumn{1}{l|}{68.90}           & \multicolumn{1}{l|}{69.33}            & \multicolumn{1}{l|}{76.39}           & \multicolumn{1}{l|}{70.80}           &73.49             & \multicolumn{1}{l|}{\textbf{92.87}}      & \multicolumn{1}{l|}{\textbf{92.25}}      & \multicolumn{1}{l|}{\textbf{92.55}}       & \multicolumn{1}{l|}{\textbf{95.04}}      & \multicolumn{1}{l|}{\textbf{92.41}}      & \textbf{93.71}       \\ \hline
\textbf{4}                                               & \multicolumn{1}{l|}{64.57}           & \multicolumn{1}{l|}{63.72}           & \multicolumn{1}{l|}{64.14}            & \multicolumn{1}{l|}{70.97}           & \multicolumn{1}{l|}{65.88}           &68.33             & \multicolumn{1}{l|}{\textbf{82.77}}      & \multicolumn{1}{l|}{\textbf{82.05}}      & \multicolumn{1}{l|}{\textbf{82.41}}       & \multicolumn{1}{l|}{\textbf{86.52}}      & \multicolumn{1}{l|}{\textbf{82.70}}      & \textbf{84.57}       \\ \hline
\textbf{5}                                               & \multicolumn{1}{l|}{52.16}           & \multicolumn{1}{l|}{51.48}           & \multicolumn{1}{l|}{51.82}            & \multicolumn{1}{l|}{59.45}           & \multicolumn{1}{l|}{54.74}           &57.00             & \multicolumn{1}{l|}{\textbf{80.98}}      & \multicolumn{1}{l|}{\textbf{79.91}}      & \multicolumn{1}{l|}{\textbf{80.44}}       & \multicolumn{1}{l|}{\textbf{87.87}}      & \multicolumn{1}{l|}{\textbf{81.19}}      & \textbf{84.44}       \\ \hline
\textbf{6}                                               & \multicolumn{1}{l|}{70.40}           & \multicolumn{1}{l|}{68.89}           & \multicolumn{1}{l|}{69.64}            & \multicolumn{1}{l|}{81.44}           & \multicolumn{1}{l|}{71.93}           & 76.39            & \multicolumn{1}{l|}{\textbf{75.41}}      & \multicolumn{1}{l|}{\textbf{74.39}}      & \multicolumn{1}{l|}{\textbf{74.90}}       & \multicolumn{1}{l|}{\textbf{82.32}}      & \multicolumn{1}{l|}{\textbf{76.02}}      & \textbf{79.05}       \\ \hline
\textbf{7}                                               & \multicolumn{1}{l|}{66.71}           & \multicolumn{1}{l|}{65.66}           & \multicolumn{1}{l|}{66.18}            & \multicolumn{1}{l|}{73.78}           & \multicolumn{1}{l|}{67.90}           &70.71             & \multicolumn{1}{l|}{\textbf{79.00}}      & \multicolumn{1}{l|}{\textbf{78.18}}      & \multicolumn{1}{l|}{\textbf{78.59}}       & \multicolumn{1}{l|}{\textbf{84.32}}      & \multicolumn{1}{l|}{\textbf{79.27}}      & \textbf{81.72}       \\ \hline
\textbf{8}                                               & \multicolumn{1}{l|}{58.93}           & \multicolumn{1}{l|}{58.34}           & \multicolumn{1}{l|}{58.63}            & \multicolumn{1}{l|}{63.63}           & \multicolumn{1}{l|}{60.19}           &61.86             & \multicolumn{1}{l|}{\textbf{81.95}}      & \multicolumn{1}{l|}{\textbf{81.38}}      & \multicolumn{1}{l|}{\textbf{81.66}}       & \multicolumn{1}{l|}{\textbf{85.59}}      & \multicolumn{1}{l|}{\textbf{82.03}}      & \textbf{83.77}       \\ \hline
\textbf{9}                                               & \multicolumn{1}{l|}{57.95}           & \multicolumn{1}{l|}{57.13}           & \multicolumn{1}{l|}{57.54}            & \multicolumn{1}{l|}{64.22}           & \multicolumn{1}{l|}{59.62}           &61.83             & \multicolumn{1}{l|}{\textbf{77.08}}      & \multicolumn{1}{l|}{\textbf{76.37}}      & \multicolumn{1}{l|}{\textbf{76.72}}       & \multicolumn{1}{l|}{\textbf{81.95}}      & \multicolumn{1}{l|}{\textbf{77.45}}      & \textbf{79.64}       \\ \hline
\end{tabular}
\label{tab: all_results}
\vspace{-3mm}
\end{table*}

\section{Results and Discussions}\label{sec:results}

\subsection{Evaluation Across Data Split Scenarios}
To comprehensively evaluate our model’s generalization capability, we conduct experiments under multiple data-split scenarios, as detailed in Table~\ref{tab:data_splitting}. 
\subsubsection{General Evaluation} In Experiment 1, we consider a merged dataset from both singers, with a random 70/20/10 train-test-validation split. 
As shown in row-1 of table~\ref{tab: all_results}, we observe a significant improvement of about 16\% with collar and 13\% without collar over the baseline, highlighting the effectiveness of our proposed TCN model. 
\subsubsection{Intra-Singer Evaluation} In Experiments 2 and 3,
we train and evaluate the model separately on each singer's recordings using distinct train-test splits. 
Rows 2 and 3 in Table~\ref{tab: all_results} show very similar or somewhat better performance as compared to Experiment-1.
This indicates that despite the added variability from combining singer styles in Experiment-1, the model generalizes well, unlike the baseline which struggles under the same conditions.
\subsubsection{Inter-Singer Evaluation}
In Experiments 4 and 5, we assess the model’s generalization across singers by training on one singer’s data and testing on the other’s. As expected, the baseline performance drops significantly in this cross-singer setup. However, our TCN-based model maintains relatively strong performance, indicating its ability to learn underlying structural patterns of ornamentations that transfer across stylistic differences.
\subsubsection{R\=ag\=a-Specific Evaluation} 
In Experiments 6 to 9, we evaluate the model’s robustness across different \textit{r\=ag\=as} by training on one raga and testing on another for the same singer. Given that ornamentation styles vary significantly across ragas, some performance drop is expected. As shown in Table~\ref{tab: all_results}, while the baseline model drops to around 60\% F1-score, our proposed model consistently maintains strong performance around 80\%, demonstrating better generalization to unseen \textit{r\=ag\=a} contexts.

\begin{figure}[t]
    \centering
    \includegraphics[scale=0.6]{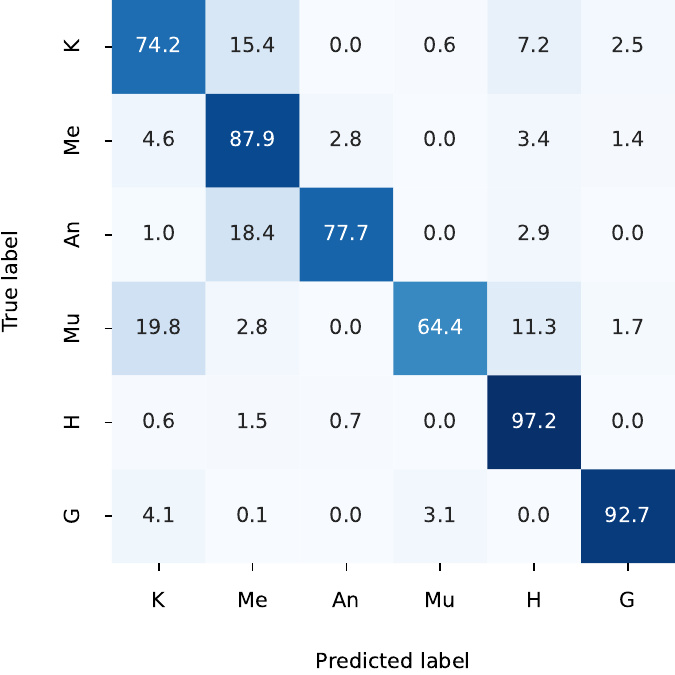}
    \caption{Confusion matrix illustrating the performance of the proposed model on the ROD dataset. The rows represent the true labels (ground truths), and the columns represent the predicted labels. K, Me, An, Mu, H, and G represent \textit{Ka\d n}, \textit{M\=ind}, \textit{Andolan}, \textit{Murki}, \textit{Ny\=as Svar}, and \textit{Gamak} respectively.}
    \label{fig:confusion_matrix}
    \vspace{-3mm}
    \end{figure}

Figure~\ref{fig:confusion_matrix} presents the confusion matrix for our best-performing configuration, which utilizes F = 120, `don't care' labelling, periodic padding, and dilated convolutions. 
The confusion matrix reveals patterns consistent with theoretical expectations and annotation trends.
For instance, \textit{Ka\d n} is often misclassified as \textit{M\=ind}, which is expected since \textit{M\=ind} is essentially a slower, more extended version of \textit{Ka\d n} articulated with a glide.
Similarly, \textit{Andolan} is frequently confused with \textit{M\=ind}, as both share a gliding character, with \textit{Andolan} essentially being a repeated or oscillatory variant of a \textit{M\=ind}.
As shown in Figure~\ref{andolan_error}, if we isolate any one glide segment within the \textit{Andolan}, it closely resembles what is labelled a \textit{M\=ind} in our dataset.
Furthermore, the model experiences significant confusion between \textit{Murk\=i} and \textit{Ka\d n}, since \textit{Murk\=i} can be conceptualized as a sequence of repeated \textit{Ka\d n svar}. Additionally, the flexible nature of \textit{Murk\=i}, by definition, complicates the classification, causing the model to perform least effectively on this class. In contrast, the model accurately classifies \textit{Ny\=as Svar} and \textit{Gamak}, due to their clearly defined structural characteristics and extended temporal duration.  

\subsection{Evaluation on concert recordings}
To evaluate the model’s generalizability to real-world Indian classical music performances, we fine-tuned our best-performing model on a set of 52 audio recordings obtained from Prasar Bharati archives. We compare it with a baseline model trained from scratch on this dataset, as well as with a model pre-trained on the ROD dataset and directly tested on the new dataset without any fine-tuning. Despite the huge domain shift, the model demonstrates satisfactory performance after minimal adaptation, as summarized in Table \ref{pb_baseline_results}.

\begin{table}[!t]
    \centering
    \renewcommand{\arraystretch}{1.3} 
    \caption{Performance comparison of the baseline and proposed models, reported with and without a 200 ms collar for the Prasar Bharati dataset. FT denotes fine-tuning}
    \vspace{2mm}
    \begin{tabular}{|c|c|c|c|c|c|c|}
        \hline
        \multirow{2}{*}{\textbf{Model}} & \multicolumn{3}{c|}{\textbf{Without Collar}} & \multicolumn{3}{c|}{\textbf{With 200 ms Collar}} \\ \cline{2-7}
        & \textbf{P} & \textbf{R} & \textbf{F1} & \textbf{P} & \textbf{R} & \textbf{F1} \\
        \hline
        \textbf{Baseline} & 53.94 & 53.41 & 53.68 & 60.83 & 56.39 & 58.52 \\
        \hline
        \makecell{\textbf{Proposed Model} \\ \textbf{(without FT)}} & 61.32 & 60.75 & 61.03 & 65.02 & 61.98 & 63.47 \\
        \hline
        \makecell{\textbf{Proposed Model} \\ \textbf{(with FT)}} & \textbf{65.26} & \textbf{64.90} & \textbf{65.80} & \textbf{69.06} & \textbf{66.18} & \textbf{67.59} \\
        \hline
    \end{tabular}
    \label{pb_baseline_results}
    \vspace{-3mm}
\end{table}

\begin{table}[!t]
\centering
\caption{Ablation Study on Proposed Model Performance for Experiment 1 ($\dagger$, PP, Di-Conv, and F denote don't care labelling, Periodic padding, Dilated Convolution, and chroma bins respectively)}
\begin{tabular}{|l|l|l|l|}
\hline
\multicolumn{1}{|c|}{\textbf{Configuration}} & \multicolumn{1}{c|}{\textbf{P}} & \multicolumn{1}{c|}{\textbf{R}} & \multicolumn{1}{c|}{\textbf{F1}} \\ \hline
\textbf{Baseline}                   & 74.92                           & 73.78                           & 74.35                            \\ \hline
\textbf{Without $\dagger$}                   & 79.55                           & 78.50                           & 79.02                            \\ \hline
\textbf{Without PP}                          & 87.82                           & 86.36                           & 86.82                            \\ \hline
\textbf{Without Di-Conv}                     & 87.61                           & 86.68                           & 87.14                            \\ \hline
\textbf{With $\dagger$, PP, and Di-Conv, F = 12}                         & 87.28                           & 86.33                           & 86.80                            \\ \hline
\textbf{With $\dagger$, PP, and Di-Conv, F = 120}     & \textbf{90.56}                  & \textbf{89.56}                  & \textbf{90.06}                   \\ \hline
\end{tabular}
\label{table:ablation}
\vspace{-3mm}
\end{table}

\subsection{Ablation studies}
\begin{figure}[t]
    \centering
    \includegraphics[height=3cm]{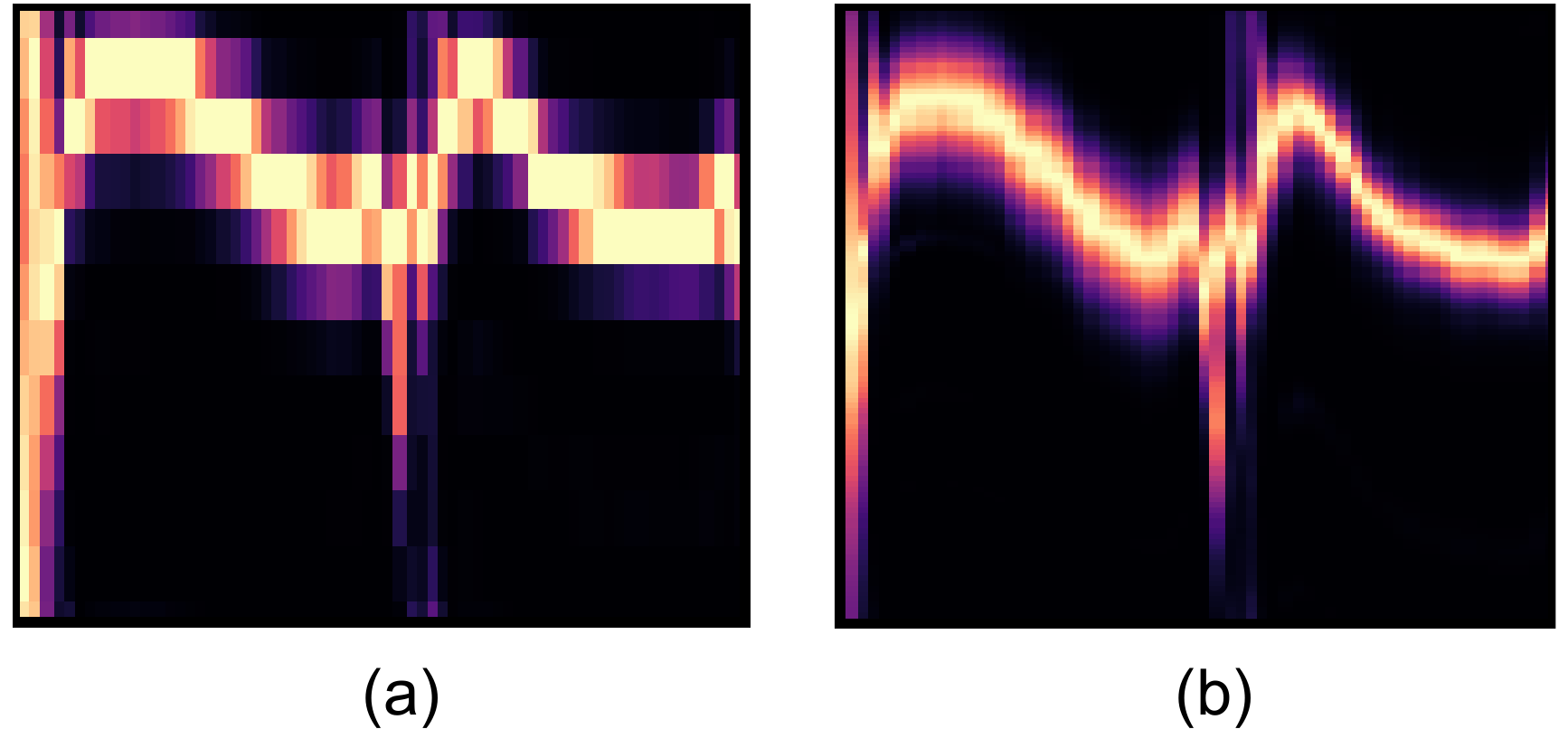}
    \caption{Comparison of (a) 12-dimensional and (b) 120-dimensional chromagrams for the same audio excerpt taken from ROD dataset. 
    }
    \label{fig:12_vs_120}
    \vspace{-4mm}
\end{figure}

Table~\ref{table:ablation} presents the results of ablation studies conducted for Experiment-1 evaluated without a collar.
The first row represents baseline CRNN model, while all others represent our proposed ED-TCN model. 
The baseline CRNN model yields an F1-score of 74.35, which is reasonable given the complexity of the ornamentation detection task. However, all configurations of our proposed model outperform the baseline significantly. Among these, we observe that increasing the chromagram resolution from the standard 12 bins to 120 further improves performance. As shown in Figure~\ref{fig:12_vs_120}, a sample for ornament \textit{Andolan} is more clearly represented with 120 bins, capturing the fine-grained, continuous pitch variations more effectively. This improvement is also reflected in the F1-score increase from 86.80 to 90.06.

We now investigate the impact of different components
in our method to identify optimal conditions for our proposed model. We study the effectiveness of `don't care labelling', the number of chroma bins, periodic padding, and dilated convolution in our
proposed TCN-based model by systematically dropping one component at a time.
The Rows 2, 3, and 4 in Table~\ref{table:ablation} show the usage of all other components except the one mentioned there. It is assumed that F = 120 unless mentioned otherwise. 

\subsubsection{`Don't care' labelling} When using both periodic padding and dilated convolutions, but without the `don't care' labelling strategy, the model still outperforms the baseline by approximately 4\%, indicating an immediate benefit of the temporal modelling framework. However, adding `don't care' labelling provides an increased performance of 11\%, which is a huge jump. In our dataset, we observe that if we do not incorporate `don't care' labeling for chunking, we end up with 31\% of the chunks containing at least one fragmented or partially labelled event. This means that the model would receive a number of misleading examples and would learn from them. 
By employing overlapping chunks and `don't care' labels, we maintain the structural integrity of audio events and exclude ambiguous regions from training, thereby facilitating more effective model training and enhancing overall performance, as is reflected in Table~\ref{table:ablation}.
It is also notable that `don't care' labelling
consistently boosts F1 scores by 7–8\%, more than either dilated convolutions or periodic padding, highlighting its effectiveness for such tasks.

\subsubsection{Periodic Padding}
To investigate the impact of padding strategies on our model's performance, we conduct experiments comparing periodic padding with conventional zero padding. 
Our results indicate a decline of more than 3\% when using zero padding as compared to periodic padding as shown in the Table \ref{table:ablation}. This degradation can be attributed to the intrinsic periodic structure of musical notes. By preserving this periodicity through periodic padding, the extracted chromagram features more accurately reflect the underlying musical patterns, thereby enhancing the model's ability to learn and generalize from the data.

\subsubsection{Dilated Convolutions} In this scenario, we evaluate the effect of dilated convolutions compared with standard convolutions. We find the performance to decrease by 3\% while using normal convolutions. This decrease is due to the coarser scale offered by dilated convolutions, enabling the model to capture a broader context and more intricate patterns in the data.

\begin{figure}[h]
    \centering
    \includegraphics[height=3cm]{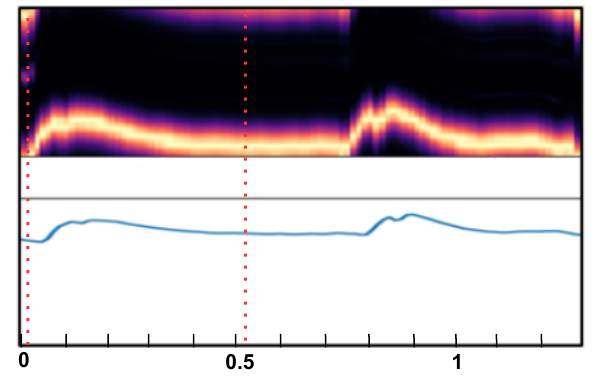}
    \caption{The figure depicts an instance of \textit{Andolan} with a cycle (marked in red lines) resembling \textit{M\=ind}. The x-axis represents time (in sec).}
    \label{andolan_error}
    \vspace{-4mm}
\end{figure}

\subsection{Subjective Analysis}
This section outlines the key sources of error in the predictions of our model and highlights the underlying complexities in the dataset and the task as follows: 
\begin{itemize}
    \item \textbf{Subjectivity in singing style across singers and \textit{R\=aga}:} A major challenge in the task of automatic ornamentation detection in IAM arises from the inherent subjectivity in singing styles, both across different singers and across different \textit{r\=agas}. Ornaments like \textit{M\=ind} or \textit{Andolan} can be executed with varying degrees of pitch deviation, duration, and subtlety, depending on the artist’s stylistic preference or the aesthetic norms of a specific \textit{r\=aga}. For instance, in the ROD dataset, we observed that in 
\textit{r\=aga} \textit{Bhairav}, the \textit{Andolan} around the notes \textit{komal re} and \textit{komal dha} is typically wide and slow, while in \textit{r\=aga} \textit{Bh\=up\=ali}, glides between the notes \textit{ga} and \textit{pa} are often very subtle and shorter. 

    \item \textbf{Inherent similarity in structure of ornaments:} Many ornamentations have similar structures inherently. For example, repeating \textit{M\=i\.nd} more than two times constitutes \textit{Andolan}. Moreover, increasing the tempo of \textit{Andolan} constructs \textit{Gamak}. One such instance is depicted in Figure \ref{andolan_error}. In this figure, the two cycles of \textit{Andolan} take on a shape similar to that of two consecutive \textit{M\=ind}. This confuses the model to predict the cycles of \textit{Andolan} as \textit{M\=ind} individually and vice versa. This intricacy is evident in Figure \ref{fig:confusion_matrix}. The red lines in Figure \ref{andolan_error} indicate the resemblance of \textit{M\=ind} with the cycles of \textit{Andolan}.
\end{itemize}

\section{Conclusion and Future Scope}

In this work, we introduce the task of automatic recognition of vocal ornamentations in Hindustani Classical Music (HCM). We present the ROD dataset containing singer-annotated HCM recordings labelled with six distinct ornamentation types and the associated \textit{r\=aga} names. Using this dataset, we develop an Encoder-Decoder Temporal Convolutional Network (ED-TCN) model that outperforms a CRNN baseline and generalizes effectively across different performers. 
Our work establishes a benchmark for automatic ornamentation detection in HCM and highlights the contribution of each model component through ablation studies.
We observe that the model benefits from periodic padding, fine-grained chromagrams, and dilated convolutions, with 'don’t care' labelling offering the strongest advantage by addressing temporal fragmentation.

The proposed ornamentation detection framework has broad applications across music pedagogy, expressive singing voice synthesis, performer identification, and content-based music recommendation. 
Future work can focus on exploring more robust domain adaptation strategies to improve model performance on real-world performance audio. 
Expanding datasets would enhance generalizability, but the annotation process remains challenging due to its high cost, intensive effort, and reliance on expert knowledge.
The subjective nature of ornament interpretation and structural overlaps among ornament types also demand hybrid approaches that combine musicological insight with data-driven learning for better performance.

\section{Acknowledgements}

We appreciate Kajal and Renu Chavan for their roles as expert vocalists and annotators. We also thank Parampreet, Sparsh, and Sharah for annotation support and Shivnarayan for managing the dataset creation. This work was supported by Prasar Bharati, India’s public broadcasting agency.

\bibliographystyle{IEEEtran}
\bibliography{references}

\end{document}